**Postural balance asymmetry and subsequent noncontact lower extremity musculoskeletal injuries among Tunisian soccer players with groin pain: A prospective case control study**

Fatma Chaari[a], Sébastien Boyas[b], Sonia Sahli[a], Thouraya Fendri[a], Mohammed A. Harrabi[a], Haithem Rebai[a], Abderrahmane Rahmani[b]

a *Research laboratory Education, Motricité, Sport et Santé, EM2S, LR19JS01, High Institute of Sport and Physical Education of Sfax, University of Sfax, Tunisia* b *Le Mans University, Movement - Interactions, Performance, MIP, EA 4334, France Faculty of Sciences and Technologies, Avenue Olivier Messiaen, 72000 Le Mans, France*

**Abstract**

**Background:** Recent studies reported postural balance disorders in patients and soccer players with groin pain (GP) compared to controls. Since postural balance asymmetry identified after an initial injury contributes for subsequent injuries, identification of this asymmetry in soccer players with GP may highlight the risk of sustaining subsequent noncontact lower extremity musculoskeletal injuries in these players. Therefore, the aims of this study were to (i) examine static and dynamic unipedal postural balance asymmetry in soccer players with GP compared to healthy ones, and (ii) quantify the risk of subsequent noncontact lower extremity injuries in these players.

**Research question:** Do soccer players with GP exhibit higher static and dynamic unipedal postural balance asymmetry, and higher risk of sustaining subsequent injuries compared to controls?

**Methods:** In this prospective case control study, 27 soccer players with non-time loss GP (GP group: GPG), and 27 healthy ones (control group: CG) were enrolled. Static and dynamic unipedal postural balance asymmetry were evaluated with a force platform using symmetry index (SI), and Y-balance test (Y-BT), respectively. Additionally, subsequent noncontact lower extremity musculoskeletal injuries were tracked for 10 months.

**Results:** The GPG revealed higher ($p<0.01$) SI in eyes closed condition, higher ($p<0.001$) side-to-side asymmetry in anterior, posteromedial and posterolateral reach distances and in composite Y-BT score compared to CG. They showed lower ($p<0.001$) composite score for injured limb and higher ($p<0.001$) side-to-side asymmetry in posteromedial reach distance compared to the cut-off values of 89.6% and 4 cm, respectively. Moreover, GPG exhibited higher odds (OR= 7.48; 95% CI= 2.15, 26.00; p<0.01) of sustaining subsequent injuries compared to CG.

**Significance:** The Y-BT should be instituted into existing pre-participation physical examinations to screen for soccer players with non-time loss GP at an elevated risk of sustaining subsequent injuries. This could help coaches and clinicians make valid return to play decisions.

INTRODUCTION

Groin pain (GP) is a complex issue [1], which encompasses pathologies from the lower abdomen, inguinal region, proximal adductors, hip joint, upper anterior thigh and perineum [2]. Different terminologies are used to describe GP including osteitis pubis, long-standing GP, pubalgia, adductor-related GP and hip joint pain [3]. To avoid confusion, the Doha agreement meeting reached consensus on a classification system to categorize GP into three major subheadings: clinical entities, hip-related GP and other causes of GP [1]. Clinical entities include adductor-, iliopsoas-, inguinal- and pubic-related GP [1]. A clinical entity could be isolated or in conjunction with other ones [4]. GP is common in sports involving directional changes [1,5], explosive movements, repeated kicking and body contact [1] as in soccer, for both amateur [6] and professional players [7]. It is a priority area for sports medicine and physiotherapy research, owing to its high prevalence, accounting for 14 to 19% of all soccer injuries [7], and propensity for recurrence [4]. This pathology could have severe consequences for the player's career, resulting in time loss or early end career [8]. Hence, GP represents a significant health and performance burden in professional soccer players [9]. Recent studies reported postural balance disorders in patients [10] and soccer players [11] with GP compared to controls.

Postural balance is the ability to maintain the center of mass over the base of support under static or dynamic conditions [12]. Postural balance during the unipedal stance is especially important because athletes drive and handle the ball with their feet by consistently using a single limb stance [13]. In soccer, the unipedal postural balance is a fundamental component of performance [14]. Its deterioration is associated with performance decrease and injuries [12]. The Y-Balance test (Y-BT) in particular, has emerged as a common screening tool to asses lower extremity injury risk in athletic populations as for instance in soccer [15]. Indeed, reduced reach distances in the anterior direction were linked to ankle sprains in a sample of both high school and collegiate American football athletes [16]. Besides, soccer players with an asymmetry ≥4 cm in the posteromedial direction are more likely to sustain a non-contact lower extremity injury [17]. Butler et al [18] found that the composite Y-BT score is predictive of non-contact lower extremity injury but no association to injury with asymmetry in any of the reach directions in collegiate American football players. Furthermore, division I collegiate athletes participating in sports including football, soccer and cross country with an asymmetry ≥4 cm in the anterior direction had a higher rate of non-contact musculoskeletal injuries [19]. However, a recent study revealed that applying previously identified between-limb symmetry thresholds (>4 cm) could not differentiate between injured and non-injured elite male youth soccer players [20]. The asymmetry identified after an initial injury contributes for subsequent injuries [21]. In this sense, a study in soccer reported further increases in subsequent injury risk for every previous injury sustained, highlighting the dangers of falling into a cycle of repeated injuries and the associated consequences, such as poor performance, deselection, or retirement [22]. Given that non–time loss injuries are associated with an elevated subsequent injury risk [23], and that many players with GP continue to train and play despite symptoms [24], identification of unipedal postural balance asymmetry in soccer players with non-time loss GP may highlight the risk of sustaining subsequent noncontact lower extremity injuries. The subsequent injuries are more severe than the initial ones owing to their greater sports incapacity (time-loss from sport) and larger burden on sports medical resources [25]. Thus, there is an urgent need to screen subsequent noncontact lower extremity musculoskeletal injuries occurrence in soccer players with non-time loss GP. Such data could help sport coaches and clinicians regarding return to play decisions, and provide them with additional information when designing training and rehabilitation programs.

Even though GP is a source of major concern in soccer players, to the best of our knowledge, little evidence is available regarding potential postural balance asymmetry or injury risk in soccer players with non-time loss GP. Therefore, the aims of the current study were to (i) examine static and dynamic unipedal postural balance asymmetry in soccer players with non-time loss GP compared to healthy ones, and (ii) quantify the risk of subsequent noncontact lower extremity musculoskeletal injuries in these players. We hypothesized that soccer players with non-time loss GP would exhibit higher (i) static and dynamic asymmetry during the unipedal posture, and (ii) incidence of sustaining subsequent noncontact lower extremity musculoskeletal injuries compared to their healthy counterparts.

METHODS

Study design

This was a prospective case control study conducted in the research laboratory, in accordance with the Declaration of Helsinki, and approved by the local Ethics Committee from the Institute "XX" (CPP, approval number: 0206/2019).

Participants

An a priori power analysis was performed with the G*Power 3.1.9.2 software to determine the convenient sample size. For a large effect size (d>0.80) [26], a power (1-β) of 0.80 and an α error =0.05, it was calculated that 26 athletes per group were required to detect significant differences between-groups in Y-BT outcomes.

Twenty-seven male soccer players with current non-time loss GP (GP group: GPG) and twenty-seven healthy ones (control group: CG) matched in terms of age, height, weight, body mass index (BMI), playing experience, weekly training and match exposure were recruited from elite soccer teams (second division). Generally, the included participants attended 4 to 5 training sessions and played one competitive match per week. The healthy soccer players were referred by their trainers or self-referral. These players were examined by a qualified physiotherapist to ensure that they had no current GP neither they suffered from a GP over the past 2 years. A qualified sports medicine physician, working on an outpatient clinic that provides health care for soccer clubs, diagnosed and referred the cases. GP diagnosis was mapped to the Doha agreement taxonomy [1]. Cases were included if they had a current unilateral GP during or after sporting activities without time loss, and classified as a clinical entity according to the Doha agreement for at least 2 months. Besides, they were eligible for inclusion if they did not benefit from any rehabilitation sessions following their current GP. The adductor-related GP and adductor-related in conjunction with inguinal-related GP were the most common entities found in our participants. However, cases with an acute onset of GP, frank inguinal hernias, pelvic bone avulsion fractures, hip joint disorders and chronic urinary tract disorder or prostatitis were excluded. The exclusion criteria for both groups were also the presence of neurological symptoms (i.e. pins and needles, tingling, and/or numbness) in the lower limbs, history of low back or sacroiliac joint pain in the past year, previous surgery in the lower abdomen, groin or the hip area, vestibular or visual disorders, musculoskeletal disease and history of injury in the past 12 months requiring medical attention and/or physical rehabilitation that could interfere in postural balance performance. Participants eligible for inclusion completed the Copenhagen Hip and Groin Outcome Score (HAGOS) questionnaire [27] to quantify groin problems at baseline and determine the severity of GP.

Procedures

*Static postural balance*

Postural balance was assessed using a force platform (PostureWin©, Techno Concept®, Cereste, France; 40Hz frequency, 12-bits A/D conversion), composed of a steel plate supported by three tri-axial transducers that records the centre of pressure (CoP) excursions. Participants were asked to stand on the force platform barefoot on one limb (non-injured [NIL] vs injured limb [IL] for the GPG, and dominant [DL] vs non-dominant limb [NDL] for the CG), and with the other limb flexed by 45° at the hip and knee. The DL was defined as the preferred limb used to kick the ball. Postural measurements were collected with the eyes opened (EO) and the eyes closed (EC). A total of 12 randomized trials were performed by each participant (3 with EO and 3 with EC for each limb) with 1 min rest periods between trials to prevent fatigue and learning effects. The mean value among the three trials performed by each participant was considered for statistical analysis.

Since the mean CoP velocity (CoP$_{Vm}$) is considered as the greatest reliable measurement among trials [28], it was selected as postural balance parameter in our study. Postural balance asymmetry between the IL and NIL (for the GPG) and DL and NDL (for the CG) were analysed using the symmetry index (SI):

$$SI = \frac{2 \times (X_{IL\ or\ NDL} - X_{NIL\ or\ DL})}{X_{IL\ or\ NDL} + X_{NIL\ or\ DL}} \times 100$$

Where *X* refers to the CoP$_{Vm}$ value. The SI is 0 if there is perfect symmetry. Positive SI values indicate higher CoP$_{Vm}$ oscillations of the IL/NDL [29].

*Dynamic postural balance*

The Y-BT, a reliable tool [30], was used to measure the unilateral dynamic postural balance in the anterior, posteromedial and posterolateral directions, and to predict lower extremity musculoskeletal injury [18]. It consists of maintaining a single limb stance, hands on waist, while reaching with the non-stance limb as far as possible along each of the three directions [30]. Participants performed the Y-BT standing on the floor at the centre of the grid with three metric tapes oriented in the three directions. The first tape was oriented in the anterior direction and was separated by 135° from the other metric tapes. These tapes were oriented on the posteromedial and posterolateral directions with 90° separating from each other. The reach distance was measured in centimetres from the centre of the grid to the point that the participant managed to reach in the specified direction. Participants performed the recommended four practice trials followed by three measured trials in each direction [30,31]. The maximum reach distance of these 3 trials in each direction was used for analysis [30]. Testing order was randomized for both limbs and 1 min rest periods were allowed between trials.

Participants' lower limb length was bilaterally measured in centimetres using a non-extensible measuring tape as the distance from the anterior superior iliac spine to the center of the ipsilateral medial malleolus. The greatest distance reached (cm) in each direction was normalised (% lower limb length) and then averaged to establish a composite score [32]. Side-to-side asymmetry, computed as the difference between both limbs (NIL or DL − IL or NDL) in each direction and in the composite score, was expressed as the absolute difference and relatively to the lower limb length. A ≥4 cm asymmetry in the posteromedial reach distance [17] and a ≤89.6% normalized composite score [18] were considered as associated with increased risk of noncontact lower extremity injury in soccer players.

*Injury surveillance protocol*

Each participant of both groups was contacted monthly by phone or e-mail for 10 months to track the first injury sustained after the initial assessments (postural balance) (Figure 1). For the purpose of this study, the injury was defined as a noncontact lower extremity musculoskeletal injury that occurred during sports participation, requiring the athlete to be absent from sports activities for at least one day [18] (e.g., muscle strain, ligament sprain, or fracture or dislocation of the foot, ankle or lower leg, knee, thigh, or hip complex, and did not include overuse injuries, contusions, and stress fractures). Injuries evaluation was carried out by a qualified sports medicine physician, and if necessary, confirmed through diagnostic imaging. Injuries that happened after the first noncontact lower extremity musculoskeletal injury during the 10 months follow-up period were not included in the analysis. Exposures were not recorded or analyzed as this information was not available at the time of contacting participants.

Statistical analysis

All statistical analyses were performed using the IBM SPSS® Statistics version 24. Data were expressed as means (M) and standard deviations (SD) or median and 25-75% interquartile range (HAGOS). Besides, the individual values of the main outcome measures in both GPG and CG were provided as supplementary materials. Normality and homogeneity of variances were assessed by the Shapiro–Wilk test and the Levene test, respectively. Independent t-tests were executed to analyse group differences for age, anthropometrics (height, weight and BMI), and normalized side-to-side asymmetry in the three directions and in the composite Y-BT scores. Besides, one-sample t-tests were executed to compare the composite Y-BT score with the cut-off value of 89.6% and the absolute side-to-side asymmetry in the posteromedial direction with the cut-off value of 4 cm in the GPG. Cohen's d effect sizes (ESs) [26] were calculated for YBT. The independent Mann-Whitney U tests were used for between-group analysis of playing experience, weekly training and match exposure, and HAGOS. A two-way ANOVA, for repeated measures, was conducted in order to analyse the effect of the group (GPG and CG) and vision (EO and EC) factors on the SI, and ESs were calculated using partial eta squared ($\eta_p^2$) [26]. For significant main factors and interactions, Bonferroni post-hoc tests were performed. Moreover, the incidence of sustaining subsequent noncontact lower extremity musculoskeletal injury was compared between the two groups using a chi-square test, and Cramer's V ESs were calculated [33]. Binary conditional logistic regression was used to assess the odds of sustaining a subsequent noncontact lower extremity musculoskeletal injury in the GPG compared to CG, with the occurrence of subsequent injury or not as dependent variable and the initial GP presence or not as the independent variable. Statistical significance was set at p<0.05.

RESULTS

Participants

Fifty-four participants were included. Descriptive characteristics of both GPG (n=27) and CG (n=27) are outlined in **Table 1**. No significant differences were found between the two groups age, height, weight, BMI, playing experience, and weekly training and match exposure. All HAGOS subscale scores were significantly lower (p<0.001) for the GPG compared to CG confirming correct group allocation **(Table 1)**.

Static postural balance

Results revealed significant main effects of the group (F=6.081; p=0.01; $\eta_p^2$=0.10) and vision (F=7.210; p=0.01; $\eta_p^2$=0.12) factors, and significant group*vision interaction (F=4.508; p=0.03; $\eta_p^2$=0.08) on the SI. The post hoc analysis revealed that the SI was significantly higher (p<0.01) in the GPG compared to the CG only in the EC condition. Vision removal significantly increased (p<0.001) the SI in the GPG **(Table 2)**. No significant difference (p=0.69) was found in the CG, when comparing EC and EO conditions.

Dynamic postural balance

Side-to-side asymmetry was significantly higher (p<0.001), in the GPG compared to the CG, in the normalized anterior (d=2.23), posteromedial (d=2.11) and posterolateral (d=1.53) reach distances as well as in the composite Y-BT score (d=2.83) **(Table 3)**.

With regard to the comparison of the composite score with the cut-off value of 89.6%, results showed significant lower (p<0.001; d=-0.84) composite Y-BT score for the IL. The side-to-side asymmetry (absolute value, cm) in the posteromedial reach distance, found in the GPG, turned out to be significantly higher (p<0.001; d=1.05) than the predictive value of 4cm.

Injury

The incidence of sustaining the first noncontact lower extremity musculoskeletal injury during the follow-up period was significantly higher (p<0.001; V=0.45) in the GPG (17/27 or 63%) compared to CG (5/27 or 18.5%). Injuries observed in our athletes are outlined in **Table 4**. The odds of sustaining subsequent noncontact lower extremity musculoskeletal injury was 7.48 times higher in the GPG compared to CG (OR= 7.48; 95% CI= 2.15, 26.00; p<0.01).

DISCUSSION

The aims of this study were to (i) examine static and dynamic unipedal postural balance asymmetry in non-time loss soccer players with GP compared to healthy ones, and (ii) quantify the risk of subsequent noncontact lower extremity musculoskeletal injuries in these players. The major results showed that soccer players with GP exhibited higher static and dynamic unipedal postural balance asymmetry compared to their healthy peers. Besides, the odds of sustaining subsequent noncontact lower extremity musculoskeletal injury was 7.48 times higher in soccer players with GP compared to controls.

Concerning static postural balance outcomes, soccer players with GP revealed higher positive SI values when vision was removed. Since these positive values indicate larger CoP oscillations of the IL [29], this suggests that soccer players with GP are dependent on visual inputs while standing on the IL. If the lower limb is moving on an unstable lumbopelvic-hip complex, structures around the groin may be stressed unduly and consequently pathologies can result [34]. Not surprisingly, one of the contributing risk factors to develop GP is poor core (known as the lumbopelvic-hip complex or trunk) stability [35]. In this context, instability of the lumbopelvic region [36] and altered trunk muscle function [1,3,10] were reported in individuals with GP compared to controls. Proprioception of the core musculature cooperate with other core musculoskeletal capacities for optimal core stability [37–40]. Deficits in proprioception has been described as both a consequence and a cause of injuries [41]. Considering that vision plays an important role to maintain postural balance, especially when another sensory receptor input is unreliable [42], and that higher visual dependence is thought to be a response to vestibular or proprioceptive impairments [43], it could be hypothesized that our soccer players with GP depended on their visual cues during the static postural balance task owing to their potential proprioception deficit of the core. Even though pelvic-proprioceptive exercises were recommended as a part of groin injury prevention programs [44], to the best of our knowledge, no data concerning proprioception impairment in athletes with GP are available.

However, in our healthy soccer players, the lack of visual feedback did not affect their static unipedal balance performance, which indicates that they are less dependent on vision during the unipedal posture. Indeed, soccer players could compensate the lack of vision by maintaining equilibrium with their proprioception abilities [45]. These authors suggested that soccer training could shift the sensorimotor dominance from vision to proprioception. Importantly, soccer players with GP performed worse compared to their healthy peers only when vision was removed. It seems that these players could not cope with the visual cues loss as the controls did. This could support our hypothesis that soccer players with GP may reveal proprioception deficit of the core.

In dynamic postural balance, our outcomes showed that soccer players with GP exhibited higher side-to-side asymmetry in the three normalized reach distances (anterior, posteromedial and posterolateral) as well as in the composite Y-BT score, when compared to controls. In contrast, Linek et al [46] reported that side-to-side asymmetry did not differ among athletes with minor hip/groin problems and controls with respect to any of the reach distances. The conflicting results could be due to the different GP categories involved in the two studies. Modifiable risk factors such as decreased hip range of motions [47], isometric adductor muscle weakness and/or a lower ratio between adductor/abductor muscle strength [48,49] and poor core stability [35] are related to an increased likelihood of suffering a GP in athletes. Lopez-Valenciano et al. [50] revealed that core stability and hip abduction isometric peak torque are associated with Y-BT performance in football players. Studies in other populations reported that the Y-BT performance is correlated with hip abductor strength and extension and external rotation of the hips [51]. Accordingly, it could be hypothesized that the modifiable risk factors to develop GP might explain the Y-BT performance in our soccer players with GP. Consequently, exploring the association between these factors and Y-BT performance in these players would be interesting to confirm or infirm such hypothesis.

It is worth noting that Y-BT is a screening tool used not only to detect dynamic postural balance disorders, but also to predict non-contact lower extremity [17,18] musculoskeletal [19] injuries. In our soccer players with GP, the absolute asymmetry in the posteromedial reach distance as well as the poor normalized composite score when compared to the cut-off values of 4cm and 89.9%, respectively, indicated that these players were at an elevated risk of noncontact lower extremity musculoskeletal injuries subsequently to GP. Similarly, it has been reported that soccer players with an asymmetry ≥4 cm in the posteromedial direction are more likely to sustain a non-contact lower extremity injury [17]. Besides, the composite Y-BT score is predictive of non-contact lower extremity injury in collegiate American football players [18]. However, no association to injury with asymmetry in any of the reach directions was reported in these players [18]. Furthermore, using the cut-off value of 4 cm, elite male soccer players were not at risk of lower extremity injuries [52]. It seems that larger asymmetries would be required to meaningfully increase the injury risk in these players. The included healthy soccer players in these studies might explain in part the conflicting results. It has been reported that soccer players with previous injuries had four to seven times greater risk for subsequent injuries [53]. Specifically, non–time loss injuries are associated with an elevated subsequent injury risk [23]. The injury surveillance outcomes in the present study confirmed that soccer players with non-time loss GP were over seven times more likely to sustain subsequent noncontact lower extremity musculoskeletal injuries compared to their healthy counterparts. However, additional studies are needed to explore the potential association between postural balance asymmetry and the occurrence of subsequent noncontact lower extremity musculoskeletal injuries in soccer players with GP.

Unfortunately, our data could not allow to specify which types of work and prevention programs must be considered to diminish the risk of subsequent injuries occurrence in soccer players with non-time loss GP. For that, determining the factors underlying the increased risk of subsequent injuries is needed to be explored in future studies to facilitate the development of these programs. Given the results of the present study, motor deficits such as the postural balance asymmetry found in soccer players with non-time loss GP should be considered while designing rehabilitation programs. Besides, the Y-BT should be instituted into existing pre-participation physical examinations to screen for soccer players with non-time loss GP at an elevated risk of sustaining subsequent injuries. This could help coaches and clinicians make valid return to play decisions.

The current study is not without its limitations. The lack of exposure rates during the follow-up period may limit the ability to generate robust logistic regression model. Thus, injury surveillance outcomes should be considered with caution. Besides, further analyses with larger sample size are needed to explore the association between postural balance asymmetry and the occurrence of subsequent injuries in soccer players with GP. Such studies could guide when designing targeted prevention programs for these players. Furthermore, assessing proprioception of the core would be of interest to confirm the hypothesis that its impairment may explain the poor static unipedal balance performance found in soccer players with GP, as compared to their peers, when vision was removed. Moreover, since postural balance in soccer players could differ among different field positions [54], analyzing postural balance outcomes in these players considering their positions could add information regarding their postural balance data.

CONCLUSION

The present study demonstrated that soccer players with non-time loss GP had static and dynamic unipedal postural balance asymmetry, and 7.48 times greater risk of sustaining subsequent noncontact lower extremity musculoskeletal injuries compared to controls. It seems interesting to consider postural balance asymmetry while designing rehabilitation programs in these players. Besides, incorporating the Y-BT into existing pre-participation physical examinations could help coaches and clinicians make valid return to play decisions regarding soccer players with non-time loss GP.